\documentclass[epj]{svjour}
\usepackage{graphics}
\usepackage{amssymb}
\usepackage{amsmath}
\usepackage{url}

\newcommand{\lambdabar}{{\mkern0.75mu\mathchar '26\mkern -9.75mu\lambda}}
\begin{document}
\title{Order-of-magnitude physics of neutron stars}
\subtitle{Estimating their properties from first principles}
\author{Andreas Reisenegger
\and Felipe S. Zepeda
}                     
%
%
\institute{Instituto de Astrof{\'\i}sica, Facultad de F{\'\i}sica, Pontificia Universidad Cat\'olica de Chile, Av. Vicu\~na Mackenna 4860, 7820436 Macul, Santiago, Chile}
\date{Received: date / Revised version: date}
%
\abstract{We use basic physics and simple mathematics accessible to advanced undergraduate students to estimate the main properties of neutron stars. We set the stage and introduce relevant concepts by discussing the properties of ``everyday'' matter on Earth, degenerate Fermi gases, white dwarfs, and scaling relations of stellar properties with polytropic equations of state. Then, we discuss various physical ingredients relevant for neutron stars and how they can be combined in order to obtain a couple of different simple estimates of their maximum mass, beyond which they would collapse, turning into black holes. Finally, we use the basic structural parameters of neutron stars to briefly discuss their rotational and electromagnetic properties.
} 
\maketitle
\section{Introduction}
\label{sec:intro}
Neutron stars (NSs) are extreme objects, with huge densities, gravitational and electromagnetic fields not encountered elsewhere, except (in some way) in black holes, but with the advantage that NSs can be directly observed. What makes them particularly fascinating for theorists is that a serious study of their properties involves all four fundamental forces of Nature (strong, electromagnetic, \\weak, and gravitational) and essentially all areas of physics: mechanics, electromagnetism, general relativity, magneto-hydrodynamics, condensed matter physics, elasticity theory, nuclear physics, quantum field theory, and probably others. Of course this list is intimidating, particularly for students just being introduced to this subject. The present article aims at lessening this feeling by using undergraduate physics to explain the most fundamental properties and estimate the numerical parameters characterizing NSs, relating them to the properties of matter in our surroundings and in white dwarfs (WDs). In this way, we intend to give a first introduction to neutron stars, transmitting a ``feeling'' for their properties and the physical basis of these. We do not aim at a comprehensive, detailed, or rigorous discussion, which can be found in various excellent textbooks, such as references \cite{Shapiro,Glendenning,Haensel}, and review articles like \cite{Lattimer}, complemented by introductory articles aimed at undergraduate students, such as \cite{Silbar,Sagert}.

Section \ref{sec:everyday} describes how quantum mechanics (through the Heisenberg uncertainty principle), together with the competition between kinetic and interaction energies, sets the main properties of atoms and nuclei and the density of condensed matter around us. In section \ref{sec:Fermi}, we discuss the equation of state (EOS) of degenerate fermion gases (based on the Pauli exclusion principle), in both the non-relativistic and ultra-relativistic limits. Section \ref{sec:WD} applies these to WDs, with strong parallels to the case of atoms, obtaining numerical estimates for the sizes, maximum mass, and escape speeds of these stars, emphasizing that a non-interacting (though possibly relativistic) fermion gas in Newtonian gravity gives physically meaningful and reasonably accurate estimates. In section \ref{sec:polytropes}, these results are reinterpreted in terms of the scaling relations obtained for polytropic EOS. The longer section \ref{sec:NS} deals with the properties of NSs. First, it is shown that beta equilibrium implies the coexistence of a majority of neutrons with a small fraction of charged particles, which do not contribute substantially to the EOS, but are crucial in stabilizing the neutrons. Then, we apply the estimate of WD sizes and maximum masses to NSs, obtaining results that are not too far from accurate calculations, but pointing out that these ignore the important effects of General Relativity and strong interactions between nuclei. Putting these in, we discuss an alternative, simple estimate of the maximum mass of NSs, proposed by Burrows and Ostriker \cite{Burrows}, assuming that the repulsion between neutrons makes the stars incompressible and imposing that their radius must remain larger than the Schwarzschild radius. We discuss how this estimate, which at face value is unphysical, because it implies an infinite speed of sound and thus violates causality, is actually quite similar to the limits obtained by imposing the causality condition. Then, we show estimates of the minimum rotation periods of various astronomical objects and discuss the concept of the ``light cylinder'' relevant for pulsars. Finally, we discuss the extremely small charge separation occurring inside NSs and other astronomical objects and the strength and consequences of NS magnetic fields.

\section{``Everyday'' matter}
\label{sec:everyday}

Even though NSs contain matter in a very extreme form, or perhaps precisely because of this reason, it is useful to start by having a look at the physics involved in the basic properties of the matter we encounter around us, on the surface of the Earth. This matter is composed of atoms, which are bound states of tiny, very dense atomic nuclei (in turn composed of protons and neutrons) surrounded by a much larger cloud of electrons. In what follows, we discuss some of their properties using simple physical estimates. A similar discussion can be found in \cite{Mahajan}.

\subsection{Size of atoms and density of condensed matter}
\label{sec:atoms}

In order to obtain the size of the electron cloud (and thus of the atom), consider the simple case of a single electron (of mass $m_e$ and charge $-e$) bound to a nucleus of charge $+Ze$. Of course we know how to solve this problem using the whole apparatus of Schr\"odinger's equation, which would give us a precise solution, but at the expense of somewhat obscuring the relevant (rather simple) physics, so we will take a much rougher approach. First of all, we note that the nucleus is much more massive than the electron, so it remains essentially motionless while the electron orbits around it. The energy of the electron, assumed to be non-relativistic (to be confirmed later), can be written as
\begin{equation}\label{electron energy}
E={\vec p^2\over 2m_e}-{Ze^2\over r}.
\end{equation}
where $\vec p$ is its momentum and $r$ its distance to the nucleus. Taking the electron wave function to have a characteristic radius $a$, to be determined, we can estimate $r\sim a$. Identifying the latter as the uncertainty in the position of the electron, $\Delta r$, the uncertainty in its momentum is bounded by Heisenberg's uncertainty principle, $\Delta p \gtrsim \hbar/\Delta r$, where $\hbar=h/(2\pi)$ is the ``reduced'' Planck constant (and $h$ the ``standard'' Planck constant). Thus, we expect $|\vec p|\equiv p\gtrsim\hbar/a$, so
\begin{equation}\label{electron energy estimate}
E\gtrsim{\hbar^2\over 2m_e a^2}-{Ze^2\over a}.
\end{equation}
In order to find the ground state, we minimize $E$ with respect to $a$, which yields the virial theorem, in the sense that the potential energy $-Ze^2/a$ must be $-2$ times the kinetic energy $\hbar^2/(2m_ea^2)$. Thus,
\begin{equation}\label{atom size}
a\sim{1\over Z}{\hbar^2\over m_ee^2}={a_0\over Z},
\end{equation}
where $a_0\equiv\hbar^2/(m_ee^2)\approx 0.5\times 10^{-8}\mathrm{cm}=0.5\ \mathrm{\AA}$ is the Bohr radius, in agreement with the much more difficult solution of the Schr\"odinger equation.\footnote{In fact, $a_0$ could be obtained in an even simpler way, as the only length that can be constructed from the three relevant dimensional constants in the problem, namely $e$, $\hbar$, and $m_e$. Of course this requires to first have the judgement to eliminate other constants, such as the speed of light and the mass of the nucleus.} From this, it is also easy to estimate the typical electron velocity,
\begin{equation}\label{electron velocity}
v={p\over m_e}\sim{\hbar\over m_ea}\sim Z{e^2\over\hbar}=Z\alpha c,
\end{equation}
where $c$ is the speed of light and
\begin{equation}\label{fine structure}
\alpha\equiv{e^2\over\hbar c}\approx{1\over 137}
\end{equation}
is the fine-structure constant, a dimensionless measure of the strength of the electromagnetic force.

This shows that the electron is non-relativistic ($v\ll c$) as long as $Z\ll\alpha^{-1}\sim 10^2$, a condition (marginally) violated by the heaviest nuclei, but adequate for the lighter ones. Note that in the opposite, ultra-relativistic limit, the electron kinetic energy is $pc\sim\hbar c/a$, with the same dependence on $a$ as the potential energy, so the total energy will no longer have a minimum. We will not dwell on this issue, whose correct treatment requires quantum electrodynamics, but we point it out because we will encounter a similar situation later in this discussion.

From eqs. (\ref{electron energy estimate}) and (\ref{atom size}), it follows that the total energy in the ground state is thus
\begin{equation}\label{Rydberg}
E_0\sim-{Ze^2\over 2a}\sim-{Z^2e^2\over 2a_0}=-Z^2\mathrm{Ry},
\end{equation}
where $1\mathrm{Ry}=13.6\mathrm{eV}=2.2\times 10^{-11}\mathrm{erg}$ is the ionization energy of the hydrogen atom. Although we had no right to expect it given our very rough approximations, the latter result is exactly the same as one obtains from solving the Schr\"odinger equation. Note that 1 eV ($=1.6\times 10^{-12}\mathrm{erg}$) is the typical thermal energy at a temperature $T\sim 1\mathrm{eV}/k\sim 10^4\mathrm{K}$, much higher than ``room temperature'', $\sim 300 \mathrm{K}$. Therefore, thermal effects can generally be ignored when studying the internal properties of atoms.\footnote{It is also interesting to note that the kinetic energies reached by protons in the Large Hadron Collider (LHC) are about $1\mathrm{TeV}\equiv 10^{12}\mathrm{eV}\sim 1\mathrm{erg}$, a macroscopic energy scale, corresponding to the kinetic energy of a marble of 2 g (thus containing $\sim 10^{24}$ nucleons) moving at $1 \mathrm{cm/s}$, whereas the highest-energy cosmic ray particles detected so far, also likely protons, have energies exceeding $\sim 10^{20}\mathrm{eV}\sim 10\mathrm{J}$, about the kinetic energy of a tennis ball in a professional match.}

When there are two or more electrons, the Pauli exclusion principle will not allow them to be in the same quantum state. This effect and their mutual repulsion (which results in an effective screening of the attraction of the nucleus on the outer electrons by the inner ones) makes the electron cloud bigger, resulting in most neutral atoms having sizes of a few $\mathrm{\AA}$ \cite{Mahajan}. Writing the radius of an atom as $fa_0$, where $f$ is a dimensionless constant in the range $\sim 1-10$, the typical mass density of ``condensed matter'', i.~e., liquids or solids, in which atoms essentially touch their neighbors (and thus are difficult to compress further), will be
\begin{equation}\label{condensed}
\rho\sim{3\over 4\pi}{Am_N\over(fa_0)^3}\sim 3.2{A\over f^3}\mathrm{g/cm^3},
\end{equation}
where $m_N$ is the mass of a nucleon (proton or neutron). This is in rough agreement with the observed densities of typical liquids and solids, $\sim 1-10\mathrm{g/cm^3}$ (recall that the gram was originally defined as the mass of $1\mathrm{cm^3}$ of water), but unfortunately the expression is very sensitive to $f$, which in turn is difficult to estimate for specific substances. For one pure substance, diamond, we know its constituents are carbon atoms ($A=12$), so its observed density of $3.5\mathrm{g/cm^3}$ would require a ``reasonable'' $f=2.2$.

\subsection{Atomic nuclei and strong interactions}
\label{sec:nuclei}

The protons and neutrons in the nucleus are held together by the strong nuclear force, which can be described as an exchange of virtual pions. Contrary to photons and gravitons (the carriers of electromagnetic and gravitational\\forces), pions are massive ($m_\pi\approx 135\mathrm{MeV}/c^2$ for neutral pions [$\pi^0$], $140\mathrm{MeV}/c^2$ for charged ones [$\pi^\pm$]), so the Heisenberg uncertainty principle allows them to move only a finite distance
\begin{equation}\label{pion}
\lambdabar\equiv{\hbar\over m_\pi c}\sim 10^{-13}\mathrm{cm}\equiv 1\mathrm{fm},
\end{equation}
where the latter unit is called \emph{fermi} or \emph{femtometer} ($10^{-15}$ m), setting the characteristic scale of nuclei, as well as their nucleon density, $n_{nuc}\sim 2\times 10^{38}\mathrm{cm^{-3}}$ (not far below $\lambdabar_\pi^{-3}$), or equivalently their mass density, $\rho\approx n_Nm_N\sim 3\times 10^{14}\mathrm{g/cm^3}$. Note that, by the same arguments given above for the electrons, any particle confined to such a small volume must have a momentum $p\gtrsim m_\pi c$, which for a nucleon of mass $m_N\approx 940\mathrm{MeV}/c^2$ (approximately the same for protons and neutrons) corresponds to a speed $v\gtrsim(m_\pi/m_N)c\sim 0.14 c$ and kinetic energy $\gtrsim(m_\pi c)^2/(2m_N)\sim 10\mathrm{MeV}$, scales we will encounter again when discussing NSs. For comparison, the electromagnetic repulsion energy between two protons at such a distance is $\sim\alpha m_\pi c^2\sim 1\mathrm{MeV}$, a relatively minor contribution to the total energy budget of the nucleus, unless there is a large number $N_p$ of protons, in which case their total repulsion energy goes up roughly as the number of proton pairs, $\sim N_p^2$.

\section{Degenerate Fermi gases}
\label{sec:Fermi}

Let us now think about WDs and NSs, which are final states of stellar evolution, whose nuclear fuel has been exhausted, and thus their thermal energy is rapidly radiated away, so their thermal pressure can no longer support them against their own gravity. However, both contain spin-$1/2$ particles (electrons and neutrons, respectively), called \emph{fermions} and subject to the Pauli exclusion principle, which states that each orbital (or one-particle quantum state) can be occupied at most by one particle. As we will see, this forces the particles to remain in motion even in the zero-temperature limit and thus provides a ``degeneracy pressure'' that does not depend on thermal effects.

\subsection{The Fermi sphere}
\label{sec:Fermi_sphere}

To determine their properties, let us start by considering the Heisenberg uncertainty principle again. From its usual form ($\Delta p \Delta r \gtrsim \hbar$) we recognize that each particle fills at least a phase-space volume of $\Delta p_x \Delta x \Delta p_y  \Delta y \Delta p_z \Delta z\sim \hbar^3$. The precise form of the latter relation gives $h^3/g_s$, since the phase-space volume $h^3$ can be occupied by particles with $g_s$ different spin projections $s_z$ along an arbitrary quantization axis. (For the spin-$1/2$ particles of interest to us, $s_z=\pm 1/2$, so $g_s=2$.) Thus, in a real-space volume $V$ and in a spherical shell in momentum space, $p<|\vec p|<p+dp$, there can be up to
\begin{equation}\label{dN}
dN={g_s\over h^3} \times V \times {4 \pi}p^2dp
\end{equation}
fermions of the same species. If we have $N$ fermions of the same kind in the lowest possible energy state (which we assume corresponds to the lowest values of $|\vec p|$, but otherwise not yet assuming a particular relation between momentum and energy), they will fill up all orbitals with $|\vec p| \leq p_F$, satisfying
\begin{equation}\label{sum of states}
N={g_s\over h^3} \times V \times \frac{4 \pi}{3} p_F^3
\end{equation}
where $V$ is the (real-space) volume occupied by the particles, the term $(4\pi/3)p_F^3$ is the volume of the ``Fermi sphere'' in momentum space, and so the whole expression is the ratio between the total phase-space volume of the system and the phase-space volume per particle. Thus, for $g_s=2$, the ``Fermi momentum'' is
\begin{equation}\label{Fermi momentum}
p_F=\left(3h^3 N\over 4\pi g_s V\right)^{1\over 3}=\hbar(3\pi^2 n)^{1/3},
\end{equation}
depending only on the particle density $n\equiv N/V$, an intensive variable, not on the extensive variables $N$ and $V$ separately. This is reassuring, as it is independent of the dimensions of the box we chose for our analysis. It is also interesting to note that, defining a typical inter-particle distance as $d\equiv n^{-1/3}$, we have $p_F=(3\pi^2)^{1/3}\hbar/d$, quite similar to the relation $p\gtrsim\hbar/a$ used in the previous section (both related to the Heisenberg uncertainty principle).

\subsection{Energy and pressure}
\label{sec:WD_energy}

In order to obtain thermodynamic quantities such as the pressure, it is convenient to consider the first law of thermodynamics, $dE=TdS-PdV+\mu dN$, where $E$ is the total (internal) energy of the thermodynamic system, $T$ its temperature, $S$ its entropy, $P$ its pressure, $V$ its volume, $\mu$ its chemical potential, and $N$ the number of particles (assumed to be all of the same species, otherwise the last term should be replaced by a sum over species, $\sum_\alpha\mu_\alpha dN_\alpha$). We are interested in the ground state, in which $T=0=S$, so we drop the first term and ignore these variables. Defining the energy density $\varepsilon=E/V$, we can write
\begin{equation}\label{pressure}
P=-\left(\partial E\over\partial V\right)_N=-\left(\partial(E/N)\over\partial(V/N)\right)_N=n^2{d\over dn} \left(\varepsilon\over n\right),
\end{equation}
a useful relation between intensive variables, valid for any single-species thermodynamic system at $T=0$.

In order to obtain the pressure of the degenerate Fermi gas, we consider its total energy, obtained by integrating over momentum using eq.~(\ref{dN}),
\begin{equation}\label{total energy}
E=\int_{p<p_F}\epsilon(p)dN={g_s\over h^3}V\times 4\pi\int_0^{p_F}\epsilon(p)p^2dp,
\end{equation}
where we are using spherical coordinates in momentum space, with a radial coordinate $p=|\vec p|$. In Special Relativity, the energy of a free particle is $\epsilon(p)=\sqrt{(mc^2)^2+(pc)^2}$, where $c$ is the speed of light. The integral can be obtained analytically (\cite{Shapiro}), but the calculation is not trivial and the result not very illuminating, so here we consider only the two limiting cases of non-relativistic ($p\ll mc$) and ultra-relativistic ($p\gg mc$) particles.

\subsection{Non-relativistic limit}
\label{sec:Non-relativistic}

In the non-relativistic limit, we expand $\epsilon(p)=mc^2+p^2/(2m)$. (We need at least these two terms, because the first, although it is much larger, corresponds to particles at rest, that do not exert any pressure.) Now, we can integrate easily and use eqs. (\ref{sum of states}) and (\ref{Fermi momentum}) to obtain
\begin{equation}\label{NR energy}
E=N\left(mc^2+{3\over 5}{p_F^2\over 2m}\right),
\end{equation}
where we identify the term $mc^2$ as the mass energy of each particle and $p_F^2/(2m)$ as the kinetic energy of the fastest moving ones. We could have guessed this relation, except for the factor $3/5$, which accounts for the fact that the kinetic energies of the particles are distributed between $0$ and $p_F^2/(2m)$. From this, we obtain the energy density
\begin{equation}\label{NR_density}
\varepsilon=mc^2n+{3^{5/3}\pi^{4/3}\over 10}{\hbar^2\over m}n^{5/3}
\end{equation}
and (using eq. [\ref{pressure}]), the pressure
\begin{equation}\label{NR_pressure}
P={(3\pi^2)^{2/3}\over 5}{\hbar^2\over m}n^{5/3}.
\end{equation}
The latter also has a simple kinetic interpretation, which can be used as a reminder or order-of-magnitude derivation: Pressure is force per unit area or momentum flux (momentum transfer per unit time per unit area). For our fermions, the typical momentum is some fraction (not much smaller than 1) of $p_F$, and their velocity is some fraction of the ``Fermi velocity'' $v_F=p_F/m$, so the pressure (momentum flux) should be some fraction of $nv_Fp_F=(3\pi^2)^{2/3}(\hbar^2/m)n^{5/3}$. Comparing to eq. (\ref{NR_pressure}), we confirm this result, also seeing that here ``some fraction'' is in fact $1/5$. Note also that, for a given number density $n$, the pressure is inversely proportional to particle mass. Thus, for a mix of different fermion species with similar abundances, the pressure is dominated by those of the lowest mass.

\subsection{Ultra-relativistic limit}
\label{sec:Ultrarelativistic}

In the ultra-relativistic limit, $\epsilon(p)=pc$. Following the same procedure, we obtain an energy per particle $E/N=(3/4)p_Fc$ (i.~e., 3/4 of the ``Fermi energy''), energy density
\begin{equation}\label{UR_density}
\varepsilon={3\over 4}(3\pi^2)^{1/3}\hbar c n^{4/3},
\end{equation}
and pressure
\begin{equation}\label{UR_pressure}
P={\varepsilon\over 3}={1\over 4}(3\pi^2)^{1/3}\hbar c n^{4/3},
\end{equation}
which again can be interpreted as a fraction of $nv_Fp_F$, but now with $v_F=c$.

It is interesting to take a look at the density at which the smooth transition between the non-relativistic and ultra-relativistic regime takes place. The condition $p_F\sim mc$ implies $n\sim (8\pi/3)\lambda^{-3}$, where $\lambda\equiv h/(mc)$ is the Compton wavelength for the relevant fermion. For electrons, $\lambda=2\pi\alpha a_0\approx 2\times 10^{-10}\mathrm{cm}$,
thus $n\sim 10^{30}\mathrm{cm}^{-3}$, whereas for nucleons (protons or neutrons), $\lambda\approx 10^{-13}\mathrm{cm}$ (about the strong interaction scale) and $n\sim 10^{40}\mathrm{cm}^{-3}$. These are easily remembered, but huge densities, as seen by comparing with the density of molecules in liquid water at ``normal'' (Earth's surface) conditions, $3\times 10^{22}\mathrm{cm}^{-3}$.

At this point, it is also interesting to briefly evaluate the assumption of zero temperature. As usual in physics, this is not an exact statement, but means that the temperature is so small that it can be ignored. In this case, it means that the thermal energies $\sim kT$ are much smaller than the typical kinetic energies of the particles, which are assumed to be due to the Pauli principle. At the transition between the non-relativistic and the ultra-relativistic regime, these energies are $\sim mc^2$, where $m$ is the mass of the relevant particles, so the condition becomes $kT\ll mc^2$. Thus, for WDs, $kT\ll m_ec^2\sim 0.5\mathrm{MeV}$, so $T\ll 10^{10}\mathrm{K}$, whereas for NSs $kT\ll m_Nc^2\sim 1\mathrm{GeV}$, so $T\ll 10^{13}\mathrm{K}$. Thus, the limits for the zero-temperature approximation are actually quite high and are amply satisfied by all but the very youngest WDs and NSs.

\section{White dwarfs}
\label{sec:WD}

\subsection{White dwarf matter and equation of state}
\label{sec:Matter}

Just as ``everyday matter'', WDs are composed of nuclei, such as He, C, O, Ne, and Mg (typically with the same number of protons and neutrons, so their ``mass number'' $A$ is twice the atomic number $Z$), as well as $Z$ electrons per nucleus. However, their typical density is much higher, and thus their inter-electron spacing $d=n_e^{-1/3}$ is much smaller than the size of an atom. We know that, for a non-relativistic electron gas of number density $n_e$, the average kinetic energy per electron is
\begin{equation}\label{kinetic-NR}
\epsilon_K^{NR}={3\over 5}{\hbar^2\over 2m_e}(3\pi^2n_e)^{2/3},
\end{equation}
whereas
the Coulomb interaction energy with the closest nucleus is
\begin{equation}\label{e-nuc}
\epsilon_{e-nuc}\sim-Ze^2\left(n_e\over Z\right)^{1/3}=-Z^{2/3}e^2n_e^{1/3}.
\end{equation}
Thus, the kinetic energy increases faster than the interaction energies as $n_e$ increases. At high densities, we reach the ultra-relativistic regime, in which
\begin{equation}\label{kinetic-UR}
\epsilon_K^{UR}={3\over 4}\hbar c(3\pi^2 n_e)^{1/3}
\end{equation}
which scales in the same way as the interaction energies. However, their ratio is $\epsilon_{e-nuc}/\epsilon_K^{UR}\sim 0.4Z^{2/3}\alpha\sim 10^{-2}(Z/6)^{2/3}$, so the interaction energies are negligible, and the electrons can be regarded as free particles, not bound to any nucleus. Over larger scales, there will be as many positive as negative charges, so their effects will cancel.

Thus, the matter in WDs can be regarded as a mix of two species of non-interacting particles: heavy, essentially motionless nuclei, which dominate the mass density,
\begin{equation}\label{ion_density}
\rho={n_e\over Z}Am_N,
\end{equation}
and low-mass, fast moving electrons, which provide most of the pressure, given by eqs. (\ref{NR_pressure}) or (\ref{UR_pressure}). Thus, in both the non-relativistic and the ultra-relativistic limit we obtain polytropic (power-law) EOSs,
\begin{equation}\label{poly_NR}
P={(3\pi^2)^{2/3}\over 5}{\hbar^2\over m_e}\left({Z\over A}{\rho\over m_N}\right)^{5/3}
\end{equation}
and
\begin{equation}\label{poly_UR}
P={(3\pi^2)^{1/3}\over 4}\hbar c\left({Z\over A}{\rho\over m_N}\right)^{4/3},
\end{equation}
respectively.

\subsection{Energy, radius, and maximum mass}
\label{sec:WD_energy}

The hydrostatic equilibrium state of these stars will be set by a balance between the gradient of this pressure and the gravitational force. As in the case of the electron wave function in the atom (see sec. \ref{sec:everyday}), we can also think of this problem as minimizing the energy of a star of fixed mass total $M$ (or electron number $N_e=ZM/A$) as a function of its radius $R$. Of course, a real star will have a certain density profile, which is characterized by more than a single parameter, but for heuristic purposes we will think of a uniform star, whose properties (for a given mass) depend only on radius.

In the non-relativistic limit, taking into account the kinetic energy of all the electrons and the gravitational binding energy of the star, the total energy is
\begin{eqnarray}\label{NR_WD_energy}
E&\sim&N_e\times{3\over 5}{p_{Fe}^2\over 2m_e}-{3\over 5}{GM^2\over R}\nonumber\\
&\sim&{3\over 10}\left(9\pi\over 4\right)^{2/3}{\hbar^2\over m_eR^2}\left(ZM\over Am_N\right)^{5/3}-{3\over 5}{GM^2\over R}.
\end{eqnarray}
Again, as in the case of the single electron in an atom (and for the same physical reasons), the kinetic energy scales with the inverse square of the radius, whereas the potential energy scales just with the inverse radius, so there is again an equilibrium radius (satisfying the virial theorem),
\begin{eqnarray}\label{WD_radius}
R_{WD}&\sim&\left({9\pi\over 4}\right)^{2/3}\left(Z\over A\right)^{5/3}{\hbar^2\over Gm_em_N^{5/3}M^{1/3}}\nonumber\\
&\sim&0.7\times 10^4\left(2Z\over A\right)^{5/3}\left(M_\odot\over M\right)^{1/3}\mathrm{km}.
\end{eqnarray}
Thus, for masses close to the solar mass, a WD will have a radius of several thousand km, not very different from that of the Earth, but with a much higher average mass density, $\rho\sim 1.3\times 10^6(A/2Z)^5(M/M_\odot)^2\mathrm{g/cm^3}$, and an electron density $n_e\sim 4\times 10^{29}(A/2Z)^4(M/M_\odot)^2\mathrm{cm^{-3}}$, roughly at the boundary where electrons become relativistic.

Contrary to planets, asteroids, and other small bodies (in which the density is roughly constant and thus $R\propto M^{1/3}$), when the mass of a WD is increased, its size decreases, and thus the density strongly increases, causing the electrons to become relativistic at large enough mass. Thus, we consider the ultra-relativistic version of eq. (\ref{NR_WD_energy}),
\begin{eqnarray}\label{UR_WD_energy}
E&\sim& N_e\times{3\over 4}p_{Fe}c-{3\over 5}{GM^2\over R}\nonumber\\
&\sim&{3\over 4}\left(9\pi\over 4\right)^{1/3}{\hbar c\over R}\left(ZM\over Am_N\right)^{4/3}-{3\over 5}{GM^2\over R}.
\end{eqnarray}
Now, both the kinetic and the potential energies are $\propto R^{-1}$, but the latter increases more strongly with $M$ than the former. The two terms are equal at a critical mass,
\begin{equation}\label{Chandra}
M_{Ch}\sim{15\over 16}(5\pi)^{1/2}\left(Z\over Am_N\right)^2m_P^3\sim 1.7\left(2Z\over A\right)^2M_\odot,
\end{equation}
where
\begin{equation}\label{Planck}
m_P\equiv\left(\hbar c\over G\right)^{1/2}=2.2\times 10^{-5}\mathrm{g}
\end{equation}
is the so-called Planck mass, the characteristic scale at which quantum gravity effects are expected. For $M<M_{Ch}$, the kinetic energy dominates, and the star will expand towards the non-relativistic regime, whereas for $M>M_{Ch}$ the gravitational energy dominates, and the star will contract, without finding a stable equilibrium as long as its constituents remain the same and no other forces come into play. Thus, $M_{Ch}$ is our estimate of the maximum mass of a WD, the so-called ``Chandrasekhar limit'', although it was first found by E. C. Stoner (see \cite{Holberg} for a nice and comprehensive discussion of the early history of white dwarf research). Since it combines quantum, relativistic, and gravitational effects (as evidenced by the presence of the constants $\hbar$, $c$, and $G$ in eq. [\ref{Planck}]), it could be called ``the first quantum gravity calculation in history''. Above this mass, no WDs can exist, and the stable equilibrium states will be either NSs or black holes.

\subsection{Escape speed}
\label{sec:WD_escape}

Since the random velocities of the electrons inside a WD are not far from the speed of light ($v_{Fe}\sim c$), one might wonder whether their escape speed is also relativistic. A direct evaluation for $M\sim M_\odot$ and $R\sim 10^4\mathrm{km}$ gives $v_{esc}=\sqrt{2GM/R}\sim 5\times 10^3\mathrm{km/s}\sim 0.02c$, thus not terribly relativistic. However, $v_{Fe}$ and $v_{esc}$ are connected by the virial theorem: The average kinetic energy per electron, $\sim(3/10)m_ev_{Fe}^2$ in the non-relativistic limit, must be $-1/2$ times the gravitational binding energy \emph{per} electron. Note, however, that the latter is \emph{not} the gravitational binding energy \emph{of} the electrons, because the gravitational binding energy is dominated by the much more massive nuclei. Thus, dropping constant factors of order unity, we require $m_ev_{Fe}^2\sim(A/Z)m_Nv_{esc}^2$, thus
\begin{equation}\label{WD_escape}
{v_{esc}\over v_{Fe}}\sim\left(Zm_e\over Am_N\right)^{1/2}\sim 0.02,
\end{equation}
consistent (in the limit $v_{Fe}\to c$) with the evaluation at the beginning of this paragraph.

\section{Newtonian stellar structure and polytropes}
\label{sec:polytropes}

\subsection{Stellar structure equations}
\label{sec:Structure}

Degenerate stars such as WDs and NSs (in addition to rocky planets and smaller bodies) are described quite well by a ``barotropic'' EOS, in which the pressure depends only on density, $P=P(\rho)$, not on other variables such as temperature (or, equivalently, specific entropy). This allows us to restrict the stellar structure equations to only two, which, for Newtonian (non-relativistic) gravity, can be written as
\begin{equation}\label{structure}
{dP\over dr}=-{Gm\rho\over r^2}, \qquad {dm\over dr}=4\pi\rho r^2,
\end{equation}
where the first gives the balance between the outward-pushing pressure gradient and the inward-pulling gravity caused by the mass $m$ enclosed within a radius $r$, and the second gives the increment in $m(r)$ as successive shells of mass density $\rho(r)$ are added. These two equations, are combined with the EOS $P=P(\rho)$, in order to calculate the stellar structure. This is done numerically by integrating outward in small steps starting from $r=0$ (where $m=0$ and the central pressure $P_c$ or density $\rho_c$ is taken as a parameter that characterizes the star) to the point $r=R$ where $P=0$, marking the stellar surface, where $m(R)=M$, the total mass of the star. Thus, one obtains the three initially unknown functions $P(r)$, $\rho(r)$, and $m(r)$ describing the structure of the star.\footnote{See \cite{Silbar} for an accessible discussion on how to implement this procedure.} In non-degenerate stars, such as, e.g. main sequence stars, energy generation and transport need to be accounted for as well, adding two additional equations for the luminosity and temperature gradients, $dL/dr$ and $dT/dr$, respectively \cite{Carroll}.

\subsection{Polytropes and scaling relations}
\label{sec:poly-scaling}

As pointed out above, the EOS of WD matter with both non-relativistic and ultra-relativistic electrons is a very special kind of barotropic EOS, namely a polytrope, $P=K\rho^\gamma$ (with the usual notation $P=K\rho^{1+1/n}$, where the constant $n$ is called ``polytropic index''). Mathematically, this is interesting, because it allows to define dimensionless functions $\hat P=P/P_c$ and $\hat\rho=\rho/\rho_c$ that, by definition, take the value $\hat P(0)=\hat\rho(0)=1$ in the center of the star, and have a unique relation (independent of $P_c$ and $\rho_c$) to each other\footnote{The readers are invited to convince themselves that this is not possible for any other functional form $P(\rho)$.} ($\hat P=\hat\rho^\gamma$). If we now also define a dimensionless radial coordinate $\hat r=r/\bar r$ and mass $\hat m=m/\bar m$, where
\begin{equation}\label{rscale}
\bar r=\left(P_c\over 4\pi G\rho_c^2\right)^{1/2}=\left(K\rho_c^{\gamma-2}\over 4\pi G\right)^{1/2}
\end{equation}
and
\begin{equation}\label{mscale}
\bar m={(P_c/G)^{3/2}\over(4\pi)^{1/2}\rho_c^2}={(K/G)^{3/2}\over(4\pi)^{1/2}}\rho_c^{(3\gamma-4)/2},
\end{equation}
the stellar structure equations take the simple form
\begin{equation}\label{dimless_structure}
{d\hat P\over d\hat r}=-{\hat m\hat P^{1/\gamma}\over\hat r^2}, \qquad {d\hat m\over d\hat r}=\hat P^{1/\gamma}\hat r^2,
\end{equation}
with a unique solution $\hat P(\hat r)$, $\hat\rho(\hat r)$, $\hat m(\hat r)$ for a given $\gamma$. In general, this solution must be found numerically, but the reader is invited to solve the special cases with $\gamma=2$ and $\gamma\to\infty$ (an incompressible fluid, with $\rho=\mathrm{constant}$ but $P$ variable), which can be done analytically.

However, regardless of the form of these specific solutions, their existence and uniqueness implies the scaling relations
\begin{equation}\label{Mscaling}
M\propto\bar m\propto\rho_c^{(3\gamma-4)/2}
\end{equation}
and
\begin{equation}\label{Rscaling}
R\propto\bar r\propto\rho_c^{(\gamma-2)/2}\propto M^{(\gamma-2)/(3\gamma-4)},
\end{equation}
from which we note various consequences:
\begin{itemize}
\item[(a)] For $\gamma=5/3$, as applicable for non-relativistic electrons, we recover the scaling $R\propto M^{-1/3}$, as in eq. (\ref{WD_radius}), and calculate the numerical factor more precisely, yielding
\begin{eqnarray}\label{poly_radius}
R_{WD}&=&4.5\left(Z\over A\right)^{5/3}{\hbar^2\over Gm_em_N^{5/3}M^{1/3}}\nonumber\\
&=&0.87\times 10^4\left(2Z\over A\right)^{5/3}\left(M_\odot\over M\right)^{1/3}\mathrm{km}.
\end{eqnarray}
\item[(b)] For $\gamma=4/3$, we see that $M$ is constant (independent of $\rho_c$), corresponding to the Chandrasekhar limit, as in eq.~(\ref{Chandra}). Again, the numerical solution yields a precise numerical value,
\begin{equation}\label{poly_mass}
M_{Ch}=3.0\left(Z\over Am_N\right)^2m_P^3=1.4\left(2Z\over A\right)^2M_\odot.
\end{equation}
Note that, in spite of the rough estimates made in the previous section, the results differed from the exact ones by only $\sim 20\%$.
\item[(c)] For all $\gamma>4/3$, $M$ is an increasing function of $\rho_c$, corresponding to stable stellar models \cite{Harrison}, whereas for $\gamma<4/3$ it is a decreasing function, corresponding to unstable and thus unphysical solutions.
\item[(d)] Among the ``physical'' solutions with $\gamma>4/3$, the harder ones ($\gamma>2$) have a radius that increases with mass or central density, whereas the opposite is true for the softer ones ($4/3<\gamma<2$).
\end{itemize}

A more general conclusion, to which we will come back later, is that, since the relations between $\rho_c$, $R$, and $M$ are always power laws, there are no ``special values'' of any of these variables, except for mathematically degenerate cases like $\gamma=4/3$ (a single mass, independent of $\rho_c$) and $\gamma=2$ (a single radius). Thus, \emph{in Newtonian gravity}, the only way to obtain a sequence of models that cuts off at a certain parameter value (maximum mass) is to have a non-polytropic EOS, in which $\gamma\equiv d\log P/d\log\rho$ is not constant. For the specific case of WDs, the EOS has $\gamma=5/3$ at low densities (implying that $M$ increases with $\rho_c$), but slowly softens to $\gamma=4/3$ at higher densities, setting an upper limit to the mass. It is important to keep this in mind for the analysis of NSs to be done below.

\section{Neutron stars}
\label{sec:NS}

\subsection{Beta equilibrium}
\label{sec:beta}

Neutrons in vacuum are unstable, decaying with a 15 minute half-life through the weak-interaction process called beta decay,
\begin{equation}\label{beta}
n\to p+e+\bar\nu_e,
\end{equation}
where $n$ denotes a neutron, $p$ a proton, $e$ an electron, and $\bar\nu_e$ an electron antineutrino, releasing
\begin{eqnarray}\label{Q}
Q&=&(m_n-m_p-m_e)c^2\nonumber \\
&=&(939.57-938.28-0.51)\mathrm{MeV}\nonumber\\
&=&0.78\mathrm{MeV}\approx 1.5 m_ec^2
\end{eqnarray}
in the form of kinetic energy of the decay products. (The masses of neutrinos and antineutrinos are small enough to be negligible, $m_\nu c^2<1\mathrm{eV}$.) On the other hand, if a proton and an electron collide with center-of-mass kinetic energy $>Q$, they can undergo ``inverse beta decay'':
\begin{equation}\label{inverse_beta}
p+e\to n+\nu_e,
\end{equation}
where $\nu_e$ is an electron neutrino. This process can happen even at zero temperature if the electron density is high enough so that the electrons are highly relativistic.

The neutrinos and antineutrinos generated in these processes are highly relativistic and very weakly interacting, therefore they will escape, making the star lose energy and approach its ground state. If the neutron, proton, and electron chemical potentials satisfy $\mu_n>\mu_p+\mu_e$, neutron beta decay is more frequent than inverse beta decay, and vice-versa for the opposite inequality, so generally a chemical equilibrium state with
\begin{equation}\label{chem_eq}
\mu_n=\mu_p+\mu_e
\end{equation}
is approached.

If thermal effects and interactions can be ignored (both of which we will discuss later), the chemical potentials are simply the Fermi energies (including the rest mass contribution), 
$\mu_i=(m_i^2c^4+p_{Fi}^2c^2)^{1/2}$. 
In the regime in which the electrons are highly relativistic ($p_{Fe}\gg m_ec$), but the protons and neutrons are still non-relativistic ($p_{Fn},p_{Fp}\ll m_Nc$), eq. (\ref{chem_eq}) becomes
\begin{equation}\label{chem_eq_1}
{p_{Fn}^2\over 2m_N}={p_{Fp}^2\over 2m_N}+p_{Fe}c.
\end{equation}
Recalling eq. (\ref{Fermi momentum}), $p_{Fi}=\hbar(3\pi^2n_i)^{1/3}$, the condition of charge neutrality, $n_p=n_e$, implies $p_{Fp}=p_{Fe}$, and thus makes the first term on the right-hand side negligible compared to the second, yielding the number density ratio
\begin{equation}\label{number_ratio}
Y\equiv {n_p\over n_n}={n_e\over n_n}\sim{n_N\over n_0},
\end{equation}
where $n_0\equiv(64\pi/3)\lambda_N^{-3}\approx 3\times 10^{40}\mathrm{cm^{-3}}$, a density at which the neutrons would already be quite relativistic ($p_{Fn}=2m_Nc$). Thus, as long as the neutrons are non-relativistic, $Y\ll 1$. The analogous derivation for the case in which all three species are relativistic is straightforward, yielding $Y=1/8$. In both cases, $n_p=n_e\ll n_n$, and correspondingly the energy density and the pressure will be dominated by the neutrons, although the presence of protons and electrons is of course crucial for the neutrons to be stable and thus present in the first place.

Up to this point, we have considered only the presence of neutrons, protons, and electrons, which is likely adequate for the lowest-density, outermost regions of the neutron star core. However, when $\mu_e>m_\mu c^2$, where $m_\mu=105.7\mathrm{MeV}/c^2$ is the mass of the muon (a much heavier lepton that otherwise shares the properties of the electron), it allows the reactions $n\to p+\mu+\bar\nu_\mu$ and $p+\mu\to n+\nu_\mu$, where $\nu_\mu$ and $\bar\nu_\mu$ are the muon neutrino and antineutrino, respectively. This allows muons to coexist in equilibrium with neutrons, protons, and electrons at somewhat higher densities. Similarly, at higher densities, other particles such as pions, kaons, hyperons, and others can appear, which are unstable in vacuum (and in nuclear physics laboratories) but, just like neutrons, are stabilized in dense matter, thanks to the Pauli exclusion principle. At progressively higher densities, the state of matter becomes increasingly uncertain, and it might even include a state of ``quark-gluon plasma'', in which all nucleons are effectively merged and quarks can move around independently. 

In what follows, we will not consider the (fairly uncertain) presence of ``exotic'' particles. In general, the presence of additional degrees of freedom reduces the neutron Fermi energy and thus the pressure, ``softening'' the EOS and thus reducing the maximum mass allowed for the stars. Therefore, the observation of a couple of stars with masses as high as $2 M_\odot$ implies that such particles, if present at all, do not play an important role in the EOS \cite{Hebeler}.

\subsection{Neutron star size and mass: a first attempt}
\label{sec:NS_first}

Thus assuming that the effects of all particles other than neutrons on the EOS are unimportant, we treat the neutron star as being composed just of neutrons. In this case, neutrons play a double role, providing both the mass (like the atomic nuclei in WDs) and the pressure (like the electrons in WDs), and one might attempt to apply the formulae derived for WDs, eqs. (\ref{poly_radius}) and (\ref{poly_mass}), putting $A=Z(=1)$ and replacing $m_e\to m_N$. The radius of a star composed of non-relativistic, non-interacting neutrons would thus be
\begin{equation}\label{NS_radius}
R_{NS}\approx 4.5{\hbar^2\over Gm_N^{8/3}M^{1/3}}=15\left(M_\odot\over M\right)^{1/3}\mathrm{km},
\end{equation}
corresponding to an average mass density $\bar\rho_{NS}\approx 1.4\times 10^{14}(M/M_\odot)^2\mathrm{g/cm^3}$ and neutron number density $n_n\approx 0.8\times 10^{38}(M/M_\odot)^2\mathrm{cm^{-3}}$, not far below the density of atomic nuclei.

Similarly, considering the limit in which the neutrons become relativistic, we can use eq.~(\ref{poly_mass}) with $A=Z=1$ to obtain a maximum mass for NSs (beyond which they would turn into black holes),
\begin{equation}\label{NS_Chandra}
M_{Ch}^{NS}\approx 3.0{m_P^3\over m_N^2}=5.7 M_\odot.
\end{equation}
One of the authors (A. R.) has for many years taught his students this result as a ``reasonable physical estimate'' of the maximum mass of NSs. Note that it relies on two key ingredients: 1) Newtonian gravity, and 2) non-interacting fermions (in this case neutrons), whose EOS becomes ``softer'' ($\gamma$ reduced to $4/3$) as they become relativistic. Neither of the two is obviously true, because, as we will discuss below, the effects of General Relativity (GR) and strong interactions are important in NSs. However, one might hope that these do not make a crucial difference, and the estimate might still be correct both qualitatively and as an order-of-magnitude estimate.

There are, in fact, a few arguments supporting it. First, it is higher than the Chandrasekhar limit for WDs (eq. [\ref{poly_mass}]) and thus allows the formation of NSs, which undoubtedly exist in the real Universe. Second, it is higher than (but of the same order of magnitude as) the largest NS masses, $\sim 2.0M_\odot$, that have been well measured from the dynamics of binary pulsars \cite{Demorest,Antoniadis}. Third, it is not enormously higher than the estimates of the maximum mass obtained from the best theoretical and observational constraints, which lie in the range $\sim 2-3M_\odot$ \cite{Hebeler}.

Of course, given that potentially important physical ingredients have been ignored, the rough quantitative agreement observed could be just coincidental. In what follows, we discuss the effects of GR and strong interactions, their implications for the maximum mass of NSs, and an alternative estimate of the maximum mass based on these two ingredients, which was suggested by Burrows and Ostriker \cite{Burrows}.

\subsection{Escape speed and Schwarzschild radius}
\label{sec:relativity}

Applying again the ``rule'' that NS properties can be obtained from WD properties through the substitutions \\$m_e\to m_N$ and $A=Z=1$, we find that in the NS case $v_{esc}$ is of the same
order of magnitude as the neutron Fermi velocity $v_{Fn}$, a result of applying the virial theorem to just one type of particles, namely the neutrons (cf. eq.~[\ref{WD_escape}]). Thus, when the neutrons become relativistic ($v_{Fn}\to c$), $v_{esc}$ also approaches $c$. This is confirmed by evaluating for ``typical'' NS parameters, $M\sim 1.5M_\odot$ and $R\sim 10\mathrm{km}$, and obtaining $v_{esc}\sim 2\times 10^5\mathrm{km/s}\sim(2/3)c$. Thus, particles orbiting near the neutron-star surface will have relativistic speeds, and GR is essential for a correct description of NS gravity.

One of the crucial concepts arising in the description of static, spherical objects (a.k.a. ``stars'') in GR is the ``Schwarzschild radius''. It can be obtained through an unreasonable extrapolation of Newtonian gravity, asking for the radius of an object whose escape speed is $c$, which yields
\begin{equation}\label{Schwarzschild}
R_S={2GM\over c^2}=3{M\over M_\odot}\mathrm{km}.
\end{equation}
Although our derivation relied on an unjustified Newtonian-relativistic hybrid, the result, when interpreted in terms of ``Schwarzschild coordinates'',\footnote{Since GR describes the space-time as curved, Euclidean geometry is not valid, and the Euclidean notion of ``radius'' can be generalized in different ways. In Schwarzschild coordinates, it is \emph{defined} as the perimeter of a circle divided by $2\pi$, but this is \emph{not} equivalent, e.~g., to a radial distance or to a radius inferred for a spherical surface from the thermal radiation received from it at a large distance.} plays a crucial role in GR, as the radius of the ``event horizon'' of a black hole, out of which no information can escape. Thus, it is also a lower limit for the radius of a star of a given mass $M$ in hydrostatic equilibrium.

We saw that, for Newtonian polytropes with $\gamma>4/3$, $M$ increases with $\rho_c$,
and there is no maximum mass. However, it is straightforward to verify that $M/R\propto\rho_c^{\gamma-1}$, so this ratio increases with $\rho_c$, meaning that at some, large enough central density, the condition $R>R_S(M)$ will no longer be satisfied. Applying this to the polytrope for non-relativistic neutrons, eq. (\ref{NS_radius}), we obtain the following upper bound for the maximum mass:
\begin{equation}\label{Sch_mass}
M_S\sim{3\pi^{1/2}\over 2^{7/4}}{m_P^3\over m_N^2}\sim 3M_\odot,
\end{equation}
nearly identical to eq.~(\ref{NS_Chandra}), except for the constant multiplying factor (which is not accurate in either case). This agreement is not accidental: In both cases we used the condition of hydrostatic equilibrium for non-interacting, degenerate neutrons (contained in eq.~[\ref{NS_radius}]). In order to obtain eq.~(\ref{NS_Chandra}), we combined it with the requirement of $v_{Fn}\to c$ (relativistic random motions of the neutrons), whereas for eq.~(\ref{Sch_mass}) we combined it with $R=R_S$ (equivalent to $v_{esc}\to c$). Thus, the virial theorem, which relates $v_{Fe}$ with $v_{esc}$ (and is contained in eq.~[\ref{NS_radius}]) actually forces the same result. This strongly suggests that, in GR, there will be a maximum mass similar to the one we already estimated, but which applies regardless of an eventual softening of the EOS.

\subsection{``Tolman-Oppenheimer-Volkoff'' (TOV) equations}
\label{sec:TOV}

In order to confirm this, we consider the stellar structure equations in GR, the so-called TOV equations, derived from the Einstein field equations (e.~g., \cite{Shapiro,Schutz}):
\begin{eqnarray}\label{TOV}
{dP\over dr}&=&-{Gm\varepsilon\over c^2r^2}\left(1+{P\over\varepsilon}\right)\left(1+{4\pi r^3P\over mc^2}\right) \left(1-{2Gm\over c^2r}\right)^{-1},\nonumber \\
{dm\over dr}&=&4\pi{\varepsilon\over c^2}r^2.
\end{eqnarray}
Comparing to their Newtonian counterpart (eq. [\ref{structure}]), we see that the mass density $\rho$ has been replaced by the energy density $\varepsilon$ (divided by $c^2$), which, in addition to the rest mass, includes the energies corresponding to random motions and (non-gravitational) inter-particle interactions. In addition, there are three correction terms in the first equation. The Newtonian form is recovered when the energy density is dominated by the rest mass ($\varepsilon\approx\rho c^2$, thus $P\ll\varepsilon\sim mc^2/r^3$) and the escape speed is much smaller than the speed of light ($2Gm/r\leq 2GM/R\ll c^2$)\footnote{All these corrections are $\sim(v_{esc}/c)^2$, thus $\sim 4\times 10^{-4}$ for a WD, and even smaller for ``normal'' stars.}. However, the general-relativistic corrections can be very important for NSs near their maximum mass, and all of them act in the direction of increasing the effective gravity (right-hand side of the first equation) with respect to the Newtonian case, thus decreasing the radius at a given mass, and decreasing the maximum mass the stars can reach. In particular, the potentially divergent term $1-2Gm/(c^2r)$ will not allow the radius of the star to become as small as the Schwarzschild radius, whereas the correction terms involving $P$ imply that the pressure has a ``weight'' that increases the effective gravity, so at large $P$ the pressure gradient can no longer prevent the collapse \cite{Harrison}.

\begin{figure}
\resizebox{0.45\textwidth}{!}{
  \includegraphics{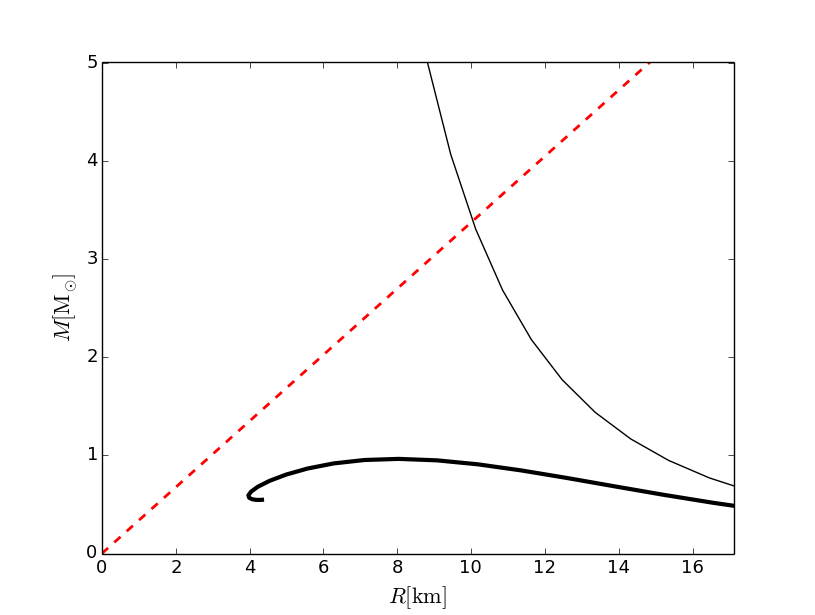}
}
\caption{Mass-radius relation for a polytropic EOS in Newtonian gravity and GR. The solid curves show the mass-radius relation for the polytropic EOS with $\gamma=5/3$ corresponding to non-relativistic, non-interacting neutrons (arbitrarily applied even in the density range where neutrons become relativistic), calculated with the stellar structure equations in Newtonian theory (eq.~[\ref{structure}]; thin line) and GR (eq.~[\ref{TOV}], thick line). For reference, the dashed line relates each mass to its Schwarzschild radius (eq.~[\ref{Schwarzschild}]). The Newtonian relation crosses the latter at $M=3.39 M_{\odot}$, $R=10.04 \mathrm{[km]}$, whereas the GR relation has a maximum at $M=0.96$, $R=8.03 \mathrm{[km]}$.}
\label{fig:Schwarzschild}       
\end{figure}

Fig.~\ref{fig:Schwarzschild} shows that, while the stellar radius is much larger than the Schwarzschild radius, the GR mass-radius relation follows the power law obtained from the Newtonian analysis. However, while nothing prevents the Newtonian relation from crossing the Schwarzschild radius at a certain value of the mass (estimated in eq.~[\ref{Sch_mass}]), the GR relation curves down before reaching it, setting a maximum mass that is lower than the previous estimate. Although the plot represents a particular polytrope, namely that of non-interacting, non-relativistic neutrons, the same qualitative behavior is observed for all polytropes with $4/3<\gamma<+\infty$.

\subsection{Strong interactions}
\label{sec:Strong}

\begin{figure}
\resizebox{0.45\textwidth}{!}{
  \includegraphics{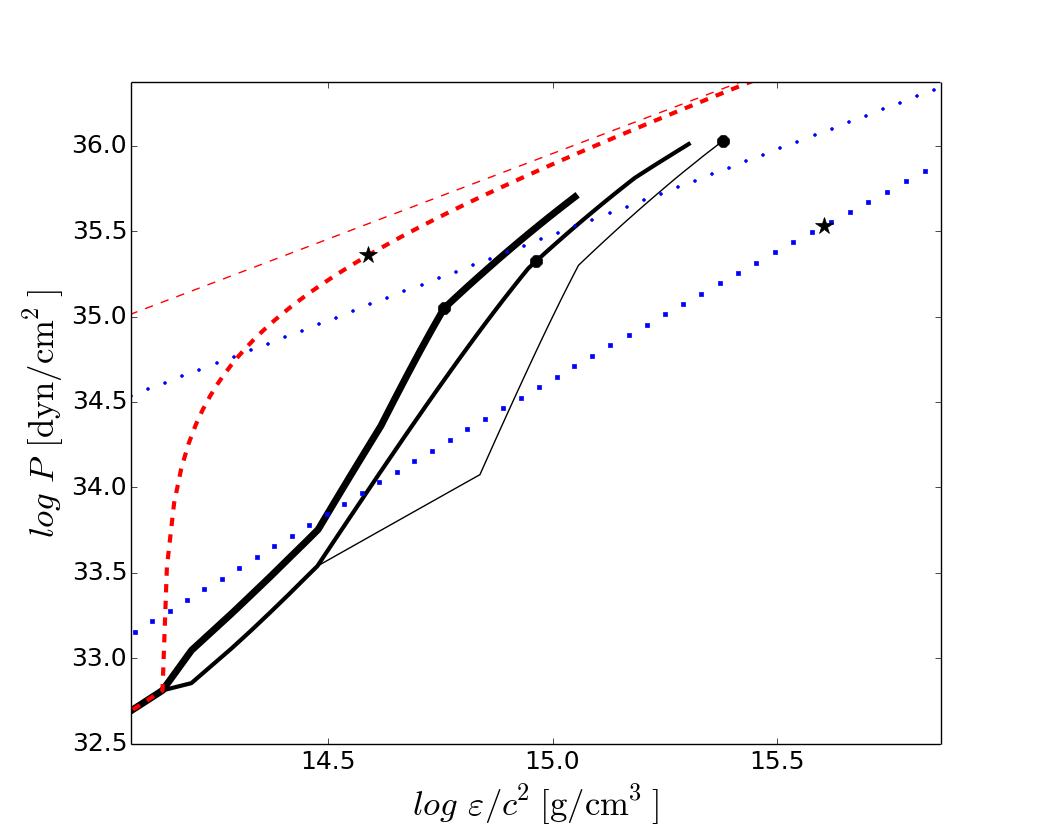}
}
\caption{EOSs for dense matter. The solid lines are three different EOS proposed by Hebeler et al. \cite{Hebeler} to cover the range allowed by the best currently available constraints from theory and observations (thick: stiff, medium: intermediate, thin: soft), with superimposed big dots representing the centers of stars of $2 M_\odot$. The other lines correspond to idealized cases: the EOS for non-interacting neutrons (thick dotted), an EOS for fully relativistic, non-interacting particles ($P=\varepsilon/3$; thin dotted), the maximum pressure allowed at a given energy density by the causality constraint ($P=\varepsilon$; thin dashed), and an EOS constructed from a causal-limit EOS ($P=P_{tr}+\varepsilon-\varepsilon_{tr}$) matched continuously to a standard BPS crustal EOS \cite{Baym71,Negele} at a transition density $\varepsilon_{tr}=1.21 \times 10^{35} \mathrm{erg/cm^3}$ (thick dashed). For these curves, the superimposed large stars represent the centers of maximum-mass stars for the respective EOSs.}
\label{fig:EOS}       
\end{figure}

Since NSs have densities around and exceeding that of atomic nuclei, strong interactions will be important. A rigorous, precise description of strong interactions is still not possible and certainly far beyond the scope of this paper. However, it is clear that the interaction between nucleons is attractive at long range, being able to hold nuclei together, and becomes increasingly repulsive at shorter range, as first suggested by Zel'dovich \cite{Zeldovich}. Thus, beyond a certain neutron density $n\sim\lambdabar_\pi^{-3}$ (see eq.~[\ref{pion}]), matter will become harder to compress than in the non-interacting case, as can be seen in Fig.~\ref{fig:EOS}, where the ``realistic'' EOSs proposed by Hebeler et al. \cite{Hebeler} stiffen around this density, contrary to the progressive softening of the EOS for non-interacting neutrons as they become more relativistic.

This motivated Burrows and Ostriker \cite{Burrows} to estimate the maximum mass of NSs by assuming that the density is set to a fixed value $\bar\varepsilon$, thus $M\sim(4\pi/3)(\bar\varepsilon/c^2)R^3$,
and imposing that $R=\beta R_S=2\beta GM/c^2$, where $\beta>1$ (plausibly $\beta\sim 2$), which yields
\begin{equation}\label{BO1}
M_{BO}=\left(3\over 32\pi\beta^3\right)^{1/2}{c^4\over G^{3/2}\bar\varepsilon^{1/2}},
\end{equation}
or, for the specific choice $\bar\varepsilon\sim m_Nc^2/\lambdabar_\pi^3$,
\begin{equation}\label{BO2}
M_{BO}\sim{(3/\pi)^{1/2}\over 2^{5/2}\beta^{3/2}}\left(m_N\over m_\pi\right)^{3/2}{m_P^{3}\over m_N^{2}}\sim 2\left(2\over\beta\right)^{3/2}M_\odot.
\end{equation}
The result is again of the same order of magnitude as the previous estimates in eqs.~(\ref{NS_Chandra}) and (\ref{Sch_mass}), however the formal expression is slightly different, including the ratio $m_N/m_\pi\sim 7$, because it is the Compton wavelength of the pions, $\lambdabar_\pi$, rather than that of the neutrons, which sets the neutron density. On the other hand, as pointed out by them, this ratio is not a large number, so in comparing this estimate with the previous ones it is overcompensated by the different (but in all cases uncertain) numerical factors. In fact, Burrows and Ostriker point out that this similarity between $m_\pi$ and $m_N$ is the reason for the similarity between the maximum masses of WDs and NSs.

It is interesting to note that the constant-energy-density case allows for an exact analytic solution of the TOV equations, the so-called Schwarzschild solution (e.~g., \cite{Schutz}), but it applies only for $\beta>9/8$, i.~e., for radii $R$ not quite as small as the Schwarzschild radius. At $\beta=9/8$, the central pressure of the star diverges, signaling that hydrostatic equilibrium is no longer possible, and the star will collapse, as discussed in depth in reference \cite{Harrison}. On the other hand, one might argue that this solution is unphysical, because the assumption of uniform density strongly contradicts the ``causality'' constraint, namely that the speed of sound, $c_s$, cannot be larger than the speed of light, $c_s^2\equiv c^2 dP/d\varepsilon\leq c^2$, since for uniform density (but non-uniform pressure, as obtained from the TOV equations) one has $dP/d\varepsilon\to\infty$. The closest we could get to this incompressible EOS would be to have $dP/d\varepsilon=1$, i.~e., $P=\varepsilon-\varepsilon_0$, where $\varepsilon_0$ is a constant. Using such a ``causal limit'' EOS everywhere in the star, the TOV equations obey scaling relations like those of the Newtonian equations with polytropic EOSs, which can be derived as an exercise or consulted in Ref.~\cite{Haensel}, Appendix E. (The same scalings also hold for the Newtonian equivalent, with $P\propto\rho-\rho_0$.) The numerical solution of the resulting dimensionless equations gives a maximum-mass star with central density only 3.03 times higher than $\varepsilon_0$, thus not extremely different from the uniform-density solution, and mass
\begin{equation}\label{CL}
M_{max}^{CL}=0.0851{c^4\over G^{3/2}\varepsilon_0^{1/2}},
\end{equation}
with the same scaling as eq.~(\ref{BO1}), and in exact agreement if one identifies $\varepsilon_0=0.24\beta^3\bar\varepsilon$.

\begin{figure}
\resizebox{0.45\textwidth}{!}{
  \includegraphics{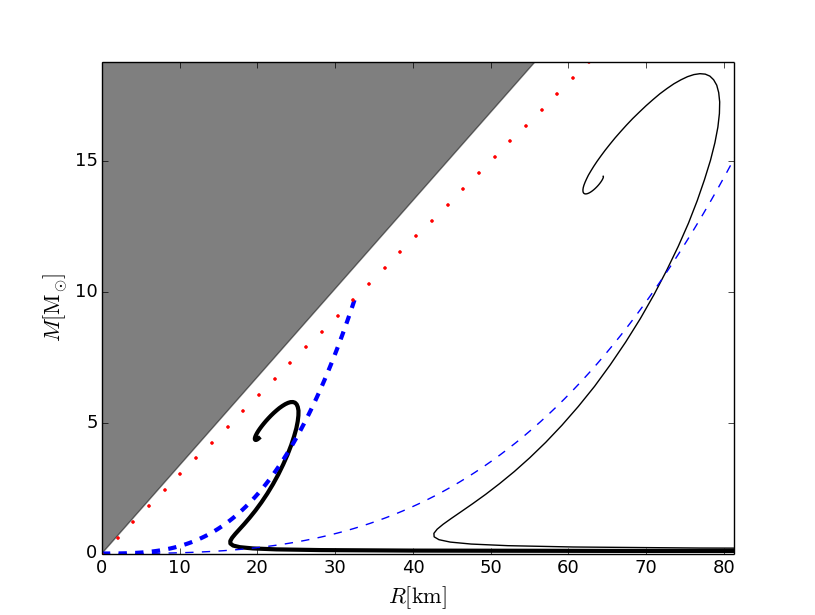}
}
\caption{Causal limit vs. Schwarzschild and constant-density models. The solid lines represent mass-radius relations for EOSs that match a realistic crustal EOS for low densities to a causal-limit EOS at high densities, with the transition occurring at the crust-core interface (thick) and at a ten times lower density (thin). For comparison, the dashed lines are mass-radius relations for uniform-density stars with densities corresponding to the respective transition densities. The dotted line corresponds to $R={9 \over 8}R_S={9 \over 4}GM/c^2$ and the grey region represents the prohibited zone due to the Schwarzschild condition.}
\label{fig:BO}       
\end{figure}

Somewhat closer to reality, one can use a well-established EOS at low densities and match it to a causal-limit EOS at some density $\varepsilon_{tr}$. Such an EOS is illustrated (together with several others) in Fig.~\ref{fig:EOS}, where one can see that, just above $\varepsilon_{tr}$, where $P\ll\varepsilon$, the effective polytropic index
\begin{equation}\label{effective_gamma}
\gamma\equiv {d\ln P\over d\ln\varepsilon}={\varepsilon\over P}{dP\over d\varepsilon}={\varepsilon\over P}\gg 1,
\end{equation}
so there is a ``wall'' that keeps $\varepsilon$ nearly constant over a substantial range of pressures, again supporting the Burrows-Ostriker estimate. Fig.~\ref{fig:BO} shows the mass-radius relations for such constructions with two different values of $\varepsilon_{tr}$, demonstrating that the mass and radius scale just as expected from that estimate, taking $\bar\varepsilon\sim\varepsilon_{tr}$.

\subsection{Rotation and light cylinder}
\label{sec:Rotation}

Once the mass $M$ and radius $R$ of a star are known, various other quantities can be calculated. Particularly interesting and historically important for NSs is the maximum rotation rate (or minimum rotation period). This is obtained by considering a matter element at the equator of a rotating star. In order for it to stay there, its gravitational acceleration must be at least as large as the centripetal acceleration. Neglecting the distortion of the star due to the rotation, this can be written in terms of the angular velocity of the star, $\Omega$, as
\begin{equation}\label{rotation}
{GM\over R^2}\gtrsim\Omega^2 R.
\end{equation}
Thus, the maximum angular velocity
\begin{equation}\label{Omega_max}
\Omega_{max}\sim\left(GM\over R^3\right)^{1/2}\sim\left({4\pi\over 3}G\bar\rho\right)^{1/2}
\end{equation}
depends only on the mean density of the star, $\bar\rho$, and the minimum period is
\begin{equation}\label{P_min}
P_{min}={2\pi\over\Omega_{max}}\sim\left(3\pi\over G\bar\rho\right)^{1/2}={1.2\times 10^4\mathrm{s}\over(\bar\rho[\mathrm{g/cm^3}])^{1/2}}.
\end{equation}
For the Earth, the Sun, a WD, and a NS, the respective densities are $\bar\rho=5.5, 1.4, \sim 10^6$, and $\sim 10^{14}\mathrm{g/cm^3}$, so their minimum periods turn out to be $P_{min}=1.4$ hours, 2.8 hours, $\sim 10 \mathrm{s}$, and $\sim 1 \mathrm{ms}$. Thus, when the first pulsars were found, with periods $\lesssim 1\mathrm{s}$, and the slowly increasing periodicity was interpreted as a stellar rotation slowing down, it was clear that the only viable physical realizations of such objects would be NSs (or even more exotic objects, such as quark stars). This conclusion was strengthened by the discovery of ``millisecond pulsars'' with periods down to $1.4\mathrm{ms}$ \cite{Backer,Hessels}. ``Sub-millisecond pulsars'', although allowed by more precise estimates of the minimum rotation period, have not yet been found.

The tangential velocity of the hypothetical matter element on the equator of the maximally rotating star is related to the escape speed, $v_{max}=\Omega_{max}R=v_{esc}/\sqrt{2}$, which is not far from $c$ in the case of NSs. This has an important consequence for the NS magnetosphere, the plasma halo around the NS coupled to it through its magnetic field. The magnetosphere is expected to co-rotate with the NS, therefore its velocity at a point at distance $r_\perp$ from the rotation axis is $v=\Omega
r_\perp$, which can of course never exceed the speed of light, and therefore the magnetosphere is limited to the so-called ``light cylinder'', of radius
\begin{equation}\label{light_cylinder}
r_\perp^{LC}={c\over\Omega}={cP\over 2\pi}=48P[\mathrm{ms}]\mathrm{km}.
\end{equation}

\subsection{Electromagnetism}
\label{sec:EM}

\subsubsection{Internal electrostatic field}
\label{sec:Electrostatic}

One little-known feature of all self-gravitating objects containing charged particles is that they must have an internal electrostatic field.

If the object contains $N_+$ particles of charge $+e$ and $N_-$ particles of charge $-e$, we will show that these two numbers must be almost exactly identical. To see this, let us assume that there is a small difference $\delta N\equiv N_+-N_-$, which can be either positive or negative. Because of Gauss' law, there will be a radial electric field $\vec E=(e\delta N/R^2)\hat r$ on the surface of the object, exerting a force $\vec F_e=\pm (e^2\delta N/R^2)\hat r$ on a charged particle of charge $\pm e$ and mass $m$ located there. For one of the two signs of charge, this force is directed outwards, and must thus be (at least) compensated by the gravitational force $\vec F_g=-(GMm/R^2)\hat r$, so the particles of this type are not expelled (which would reduce $|\delta N|$). This requires
\begin{equation}\label{Imbalance}
|\delta N|\leq {GMm\over e^2}\sim{Gm_Nm\over e^2}N_N\sim 10^{-36}{m\over m_N}N_N,
\end{equation}
where $N_N=M/m_N$ is the total number of nucleons in the object. In a NS, $N_N$ is roughly the same as the number of neutrons, and perhaps $\sim 10-100$ times larger than the number of charge carriers, $N_+$ or $N_-$. Thus, because of the much smaller magnitude of gravitational forces between charged particles compared to their electrostatic interactions ($Gm_N^2/e^2\sim 10^{-36}$), the numbers of positive and negative charge carriers must be identical to at least 34 significant figures in order to prevent charged particles to be ejected from a NS.

Inside a NS, since the proton and electron densities are similar, but their masses are very different, the electrons are subject to a much larger degeneracy pressure, whose radial gradient pushes them outwards, whereas the protons feel a much larger gravitational force, pulling them inwards. The only
way to hold the two species in equilibrium is through a slight charge separation (a small excess of protons closer to the center) that creates an outward-directed electric field that pushes protons outwards (balancing their gravitational force) and pulls the electrons inwards (balancing their pressure gradient). From the former condition, and following essentially the same derivation as in the previous paragraph, now applied to a sphere of radius $r<R$, we obtain $\delta N(r)/N_N(r)\approx Gm_N^2/e^2\sim 10^{-36}$, i.~e., there \emph{must} be a tiny charge imbalance, causing a radial electric field such that its outward force on the protons is roughly the same as their weight (in fact slightly smaller, because we have ignored the small correction from the proton pressure gradient). This effect has been considered, e.~g., in Ref. \cite{Reisenegger06}, but we are not aware of any important observational consequences. Also, the readers are invited to estimate the magnetic field produced by rotating this charge distribution and verify that it is much smaller than the magnetic fields observed in NSs, which we discuss next.

\subsubsection{Magnetic field}
\label{sec:B}

As far as we know, and contrary to nearly everything said so far in this article, the magnetic field of NSs cannot be inferred from fundamental physical principles, but likely depends on the formation history of the particular star. The reason is that NS interiors, like most astrophysical systems, can be regarded as highly conducting plasmas, in which the magnetic field and its supporting currents evolve only on very long time scales (e.~g., \cite{Baym69,GR92,Reisenegger09}).

However, since the magnetic field has a positive energy density $B^2/(8\pi)$, and a hypothetical expansion of a star conserves the magnetic flux, $BR^2=\mathrm{constant}$, such an expansion will reduce the total magnetic energy,
\begin{equation}\label{E_B_scaling}
E_B\sim {B^2\over 8\pi}\times{4\pi R^3\over 3}={B^2 R^3\over 6}\propto{1\over R}.
\end{equation}
Thus, this expansion will occur as long as it is not prevented by the gravitational force. Therefore, a firm (and likely very conservative) upper bound on the (root-mean-square) magnetic field in any star can be obtained by requiring that $E_B$ is smaller than the absolute value of the gravitational binding energy, resulting in the condition
\begin{equation}\label{B_max}
B\lesssim\left({18\over 5}{GM^2\over R^4}\right)^{1/2}\sim 10^{18}{M/M_\odot\over(R/10\mathrm{km})^2}\mathrm{G}.
\end{equation}
More stringent upper bounds on the magnitude, as well as conditions on the geometry of the magnetic field, can be obtained (so far non-rigorously) by analyzing the hydromagnetic stability of the magnetized stellar fluid (e.~g., \cite{Braithwaite09,Reisenegger09}). The dipole components of NS magnetic fields inferred from their spin-down (see below) range from $\sim 10^8\mathrm{G}$ for millisecond pulsars through $\sim 10^{12}\mathrm{G}$ for ``average'' classical pulsars up to $\sim 10^{15}\mathrm{G}$ for magnetars, still much lower than the upper limit. Stronger toroidal fields might be present inside the NSs. For comparison, the strongest magnetic fields ever produced by humans (and only for a few microseconds) are $\sim 10^7\mathrm{G}$.

As mentioned above, the charge density inside NSs is much too small for its solid-body rotation to produce the observed fields. On the other hand, the energy cost of aligning the spins of the degenerate particles, given the constraints set by the Pauli exclusion principle, is also quite prohibitive. Thus, the magnetic field must be supported by a current density $\vec j$ due to a relative velocity $\vec v_{rel}$ between positive and negative charges (taken to have charges $\pm e$ and number densities $n_p=n_e$). Using Amp\`ere's law (and neglecting the ``displacement current'' term, because we are considering an essentially static electromagnetic field), we have
\begin{equation}\label{current}
{c\over 4\pi}\nabla\times\vec B=\vec j=n_ee\vec v_{rel}.
\end{equation}
Thus, if we assume that $\vec B$ varies on a spatial scale $\sim R$ and thus estimate $|\nabla\times\vec B|\sim B/R$, we have
\begin{equation}\label{relative}
v_{rel}\sim{cB\over 4\pi n_e e R}\sim 5\times 10^{-12}{B_{12}\over n_{36}R_6}\mathrm{cm/s},
\end{equation}
i.~e., due to the huge density of charge carriers the required relative velocity is so small that it would take two given, opposite charges $\sim 10^{10}n_{36}R_6^2/B_{12}$ years to move away from each other by a distance comparable to one neutron-star radius.

\section{Conclusions}
\label{sec:Conclusions}

We hope to have been able to give students a first glimpse at the very extreme properties of neutron stars, showing how they can nearly all be understood in terms of basic physical principles. We believe that this can give them a firm grounding to study the vast specialized literature on the subject and to put more specific and precise result in context.

\section{Acknowledgements}

We thank S. Guillot and two anonymous referees for useful comments that improved this article. This work is partly based on the undergraduate research project in astronomy (Pr\'actica de Licenciatura) carried out by FSZ under the supervision of AR. It was funded by FONDECYT Regular Projects 1110213 and 1150411, CONICYT International Collaboration Grant DFG-06, and CONICYT Basal Funding Grant PFB-06.

\end{document}